# Wide Area Network Intelligence with Application to Multimedia Service

Satoshi Kamo, Yiqiang Sheng
Chinese Academy of Sciences, Beijing, China
University of Chinese Academy of Sciences, Beijing, China
Graduate School of Engineering, Osaka University, Osaka, Japan
Department of Communications and Integrated Systems, Tokyo Institute of Technology, Tokyo, Japan
E-mail: shengyq@dsp.ac.cn

*Abstract*—Network intelligence is a discipline that builds on the capabilities of network systems to act intelligently by the usage of network resources for delivering high-quality services in a changing environment. Wide area network intelligence is a class of network intelligence in wide area network which covers the core and the edge of Internet. In this paper, we propose a system based on machine learning for wide area network intelligence. The whole system consists of a core machine for pre-training and many terminal machines to accomplish faster responses. Each machine is one of dual-hemisphere models which are made of left and right hemispheres. The left hemisphere is used to improve latency by terminal response and the right hemisphere is used to improve communication by data generation. In an application on multimedia service, the proposed model is superior to the latest deep feed forward neural network in the data center with respect to the accuracy, latency and communication. Evaluation shows scalable improvement with regard to the number of terminal machines. Evaluation also shows the cost of improvement is longer learning time.

*Index Terms*—machine learning, wide area network, terminal-related systems, multimedia service.

## 1. Introduction

The current models of machine learning are almost limited in a data center or local area network due to the challenge of communication. The average latency and the communication cost between the data center and the widely distributed users are two of critical problems to be solved. As a potential solution, wide area network (WAN) intelligence extends the existing distributed models of machine learning for data processing in a data center or local area network to the neighborhood of the widely distributed users.

Network intelligence [1] is a discipline that builds on the capabilities of network systems to act intelligently by the usage of network resources for delivering high-quality services in a changing environment. The WAN intelligence is a class of network intelligence for geographically distributed data processing in wide area network which covers the core and the edge of Internet. Because the terminal computational resources of the WAN intelligence are set in the geographical neighborhood of the end users, the communication cost between the end users and terminal machines is naturally trivial, as shown in Fig. 1. That makes it easy to process geo-distributed data efficiently and accomplish faster terminal response to the end users.

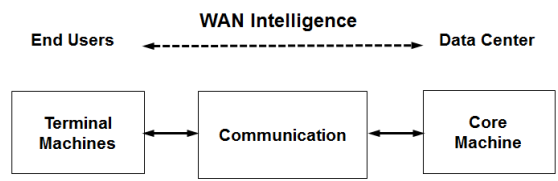

**Fig. 1** Wide area network (WAN) intelligence makes it easy to process the distributed data and provide terminal responses.

With the rapid growth of information networking technologies, many terminal-related techniques such as P2P networks [2], wireless sensor networks [3] and the cache system [4] are generating huge amounts of data which are distributed on each corner of Internet. The data are valuable to improve the quality of service or the accuracy of prediction, but it is still a big challenge to process distributed data efficiently. Cloud computing partly meets the challenge by collecting data and resources together. Since most data from terminals are naturally distributed, the current cloud computing is inefficient because the communication overheads to collect the data are high. In fact, most distributed models of machine learning are only applicable to local area network due to the same reason. To improve the efficiency of cloud computing, many techniques including geo-distributed data processing [5], fog computing, grid computing, and parallel algorithms have been discussed.

The contributions of this paper are as follows. (1) We propose a dual-hemisphere model of machine learning for the WAN intelligence to accomplish faster terminal response for the requests from the end users. (2) We investigate the architecture and the procedure to realize the WAN intelligence with application to multimedia service. (3) Evaluations show the latency, the accuracy and the communication cost are greatly improved. (4) Evaluations also show the cost of above improvement is longer learning time.

The rest of this paper is organized as follows. Section 2 provides the related work. Section 3 is the architecture. Section 4 presents the dual hemispheres. Section 5 shows the procedure. Section 6 implements the evaluation. Section 7 discusses the proposal. Section 8 concludes this research.

## 2. Related Work

O. Kurasova *et al.* [6] presented an overview of methods and technologies used for big data clustering. Neural network-based self-organizing maps and their extensions for big data clustering were reviewed. M. Malleswaran *et al.* [7] investigated various neural networks including bidirectional associative memory neural network for global positioning system and inertial navigation system data integration. K. A. Jadda *et al.* [8] proposed a scalable probabilistic graphical model to overcome these limitations of Bayesian networks for massive hierarchical data. They applied the model to solve two challenging probabilistic-based problems on massive hierarchical data sets for different domains, namely, bioinformatics and latent semantic discovery over search logs.

I. Stojmenovic *et al.* [9] discussed Fog computing which extended Cloud computing. The research elaborated the motivation and advantages of Fog computing, and analyzed its applications in a series of real scenarios, such as Smart Grid, smart traffic lights in vehicular networks and software defined networks. D. Strigl *et al.* [10] presented the implementation of a framework for accelerating the training and classification of convolutional neural networks on the GPU. Experiment showed that the training and classification on GPU performs 2 to 24 times faster than on CPU depending on the network topology. F. Ino [11] introduced a cycle sharing system which exploited idle GPU cycles for acceleration of scientific applications. It made GPU useful not only for graphics applications but also for general applications. L. Bako *et al.* [12] implemented a Spiking Neural Network on a CUDA driven Nvidia video-card. The results showed the computation speed and the classification accuracy were improved. Y. Tsuchida *et al.* [13] proposed a parallelized method for calculation of back propagation by GPU. The results showed the proposed method is 25 times faster than the non-parallelized method. B. Tomislav *et al.* [14] proposed a parallel implementation of feed forward neural network on GPU, and the results showed the effectiveness of the proposed parallel algorithm.

However, most researches are limited in Cloud-based services and GPU-based algorithms. It is not enough to apply them to wide area network because many serious problems are not sufficiently discussed. For example, the latency and the communication are two of the most important problems for terminal-related systems.

## 3. Architecture

The motivation is the calculating hemispheres [15] from the studies of split-brain patients. In general, a whole brain consists of two hemispheres with a corpus callosum connecting them. If the right and left hemispheres are separated without the communication through the corpus callosum, each hemisphere may have separate perception, concepts, and impulses to act. It means that it might be possible to pre-train the right and left hemispheres separately in different geo-distributed places using different data sets. Then, the two hemispheres are fine-tuned through an interaction just like the function of the corpus callosum to avoid the conflict between the right and left hemispheres.

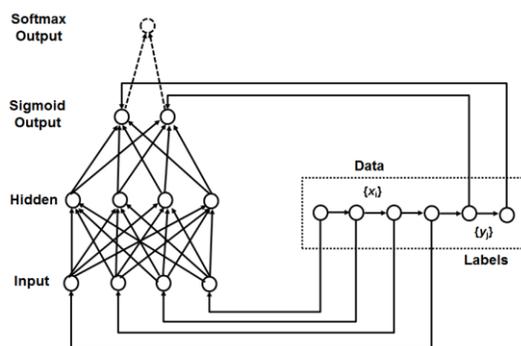

**Fig. 2** A deep feed forward neural network (DFFNN) is trained after collecting all geo-distributed data in the data center.

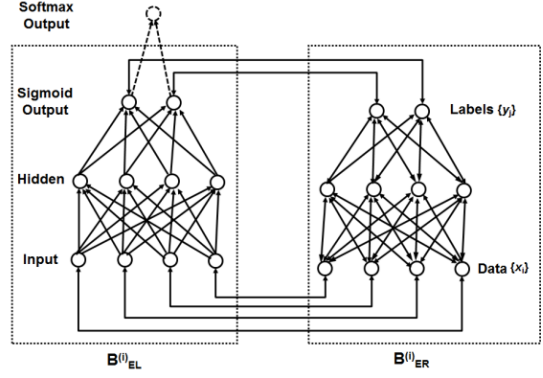

**Fig. 3** The dual-hemisphere model which consists of right and left hemispheres is trained without the collection of geo-distributed data.

For comparison, deep feed forward neural network (DFFNN) is trained by using a data set $\{x_i\}$ as the input with a labels set $\{y_j\}$ as the output in the data center, as shown in Fig. 2. The centralized data set in the data center is marked as $D_c$. It is expensive to process geo-distributed data, since it takes much overhead to collect all distributed data from different places. Furthermore, it is a big challenge to provide a fast terminal response of the request from end users, because of the long latency between the DFFNN in the data center and the widely distributed users. The centralized data are relatively large-scale, while the distributed data are with some personalized features based on different users. It is expensive to transmit the data between the users and the data center. Besides, the personalized features of the distributed data might degrade the accuracy of model, since all data are proceeded uniformly using the DFFNN model in the data center.

In order to process distributed data more efficiently and accomplish shorter latency, the architecture of the proposal consists of a core machine and many terminal machines. The terminal machine is in the geographical neighborhood of the end users. Each terminal machine has a dual-hemisphere model which consists of a right hemisphere $B^{(i)}_{ER}$ for data generation and a left hemisphere $B^{(i)}_{EL}$ for terminal responses, as shown in Fig. 3, where $1 \leq i \leq m$ and $m$ is the number of terminal machines. As a result, the communication between the terminal machines and the corresponding end-users is trivial, and the main cost of communication is between the core machine and the terminal machines.

The initial parameters of the left hemisphere $B^{(i)}_{EL}$ and the right hemisphere $B^{(i)}_{ER}$ are downloaded from the core machine with a global dual-hemisphere model which consists of a left hemisphere $B_{CL}$ and a right hemisphere $B_{CR}$ pre-trained by using the centralized data set $D_c$. In the terminal machine, a training loop between the left hemisphere $B^{(i)}_{EL}$ and the right hemisphere $B^{(i)}_{ER}$ is designed to keep learning using the distributed data set $D_d$. For the dual-hemisphere models, the difference between the core machine and the terminal machines is the data sets, i.e. $D_c$ for the core machine and $D_d$ for the terminal machines. As a result, the communication between the terminal machines and the core machine is only using the parameters of the dual-hemisphere models. That is to say, it is not necessary to transmit any original data.

## 4. Dual Hemispheres

In practice, a unidirectional neural network is used to be the left hemisphere, and a bidirectional neural network is used to be the right hemisphere. The data set is marked as $D$ including the centralized data set $D_c$ and the distributed data set $D_d$. For both the data sets, the input data are denoted as $\{x_i\}$ and the output labels are marked as $\{y_j\}$. For the unidirectional neural network, as shown in Fig. 2 and the left hemisphere of Fig. 3, the input of each neuron is as the following vector.

$$X_j = (x_1, x_2, ..., x_{d_j})^T.$$

where $d_j \leq d_{max}$ is the in-degree of the $j^{th}$ neuron, and $d_{max}$ is the maximum degree of neurons. If the output of each neuron is marked as $y_j$, then the unidirectional mapping is satisfied as the following.

$$y_j = f(W_j X_j + b_j).$$

where $f$ is the activation function, $b_j$ is the bias of the $j^{th}$ neuron, $W_j = (w_{1j}, w_{2j}, ..., w_{ij}, ...)$ is the vector of weights, $w_{ij}$ is the weight from neuron $i$ to neuron $j$, and $i = 1, 2, ..., d_j$. The feed forward neural network belongs to the above unidirectional neural network. The sigmoid function is used to be the activation function, then the following mapping is satisfied.

$$y_j = \frac{1}{1+\exp(W_j X_j + b_j)}.$$

For the softmax output layer, the following function is used as the output of the machine learning system.

$$soft\max(y_j) = \frac{\exp(y_j)}{\sum_i \exp(y_i)}.$$

For the bidirectional neural network, as shown in the right hemisphere of Fig. 3, the positive direction from $\{x_i\}$ to $\{y_j\}$ is the same as the unidirectional neural network. For the negative direction from $\{y_j\}$ to $\{x_i\}$, the input of each neuron is as the following.

$$Y_i = (y_1, y_2, ..., y_{d_i})^T.$$

where $d_i \leq d_{max}$ is the in-degree of the $i^{th}$ neuron in the negative direction from $\{y_j\}$ to $\{x_i\}$, and $d_{max}$ is the maximum degree of neurons. If the output of each neuron is marked as $x_i$, the mapping is satisfied.

$$x_i = g(W_i Y_i + b_i).$$

where $g$ is the activation function in the negative direction from $\{y_j\}$ to $\{x_i\}$, $b_i$ is the bias of the $i^{th}$ neuron, $W_i = (w_{1i}, w_{2i}, ..., w_{ji}, ...)$ is the vector of weights, $w_{ji}$ is the weight from neuron $j$ to neuron $i$, and $j = 1, 2, ..., d_i$. If the same sigmoid function is used to be the activation function in the negative direction from $\{y_j\}$ to $\{x_i\}$, then the mapping is satisfied.

$$x_i = \frac{1}{1+\exp(W_i Y_i + b_i)}.$$

The bidirectional neural network is used to be the right hemisphere for data generation. The input is a given label in the labels set $\{y_j\}$. The output is the generated data marked as $x$. The unidirectional neural network is used to be the left hemisphere for terminal response. The input is the generated data $x$. The output is the response marked as $y$ to the end-users.

The training error with respect to the set of labels $\{y_j\}$ is defined as the following.

$$E^t_{(1)} = \frac{1}{2} \sum_j \|y - y_j\|^2.$$

The generation error with respect to the set of unlabeled data $\{x_i\}$ is defined as the following.

$$E^r_{(2)} = \frac{1}{2} \sum_i \|x - x_i\|^2.$$

Let $\theta = \{\theta_i\}$ designate the set of all parameters including $\{W_i\}$ and $\{b_i\}$. For calculating the regularization error, Lasso and Ridge terms are defined by the $q$-norm as follows.

$$R_q = \|\theta\|_q = \left(\sum_i |\theta_i|^q\right)^{1/q}.$$

where $q$ is an integer, $q = 1$ is for the Manhattan norm and $q = 2$ is for the Euclidean norm.

Then, the cost function including the training error, the generation error and the regularization error is defined as the following.

$$E(\theta, D)_{(3)} = E^t + E^r + \sum_q \lambda_q R_q.$$

where $\theta$ is the set of all parameters, $D$ is the set of all data, $E^t$ is the training error with respect to the set of labeled data $\{x_j, y_j\}$, $E^r$ is the generation error with respect to the set of unlabeled data $\{x_i\}$, $R_1$ is the Lasso regularization term with a weight $\lambda_1$, and $R_2$ is the Ridge regularization term with a weight $\lambda_2$. The parameter optimization is as the following.

$$\theta^*_{(4)} = \arg \min_\theta E(\theta, D).$$

where $\theta^*$ is the set of optimized parameters.

The procedure of optimization is based on the equations (4) to construct the dual-hemisphere model including a unidirectional neural network and a bidirectional neural network. The target of the right hemisphere is to minimize the generation error with inputting a label $y_j$, while the target of the left hemisphere is to minimize the cost function with inputting a datum $x_i$.

## 5. Procedures

In order to realize the wide area network intelligence, the main procedure to construct the dual-hemisphere model of machine learning is based on the equations (1), (2), (3) and (4) as the following.

---

Procedure (I): The construction of the dual-hemisphere model of machine learning.
Input: The data set $D$, a unidirectional neural network and a bidirectional neural network with the set of parameters $\theta$.
Output: An optimized model of neural networks with the set of optimized parameters $\theta^*$.

Step 1: Initiation: A unidirectional neural network pre-trained by procedure (II) based on the centralized data set $D_c$ is used to be the initial global left hemisphere $B_{CL}$; A bidirectional neural network pre-trained by procedure (III) based on the distributed data set $D_c$ is used to be the initial global right hemisphere $B_{CR}$; The initial parameters of all local left hemispheres $\{B^{(i)}_{EL}\}$ are downloaded from the initial global left hemisphere $B_{CL}$; The initial parameters of all local right hemispheres $\{B^{(i)}_{ER}\}$ are downloaded from the initial global right hemisphere $B_{CR}$.

Step 2: Getting the distributed data set $D_d$ to minimize the training error of the local left hemispheres $\{B^{(i)}_{EL}\}$ in the gradient descent way, i.e. $\Delta \theta_i = -\lambda_0 \partial E^t / \partial \theta_i$, where $\lambda_0$ is a user-defined non-zero small number and the default value is 0.01.

Step 3: Getting the distributed data set $D_d$ to minimize the generation error of the local right hemispheres $\{B^{(i)}_{ER}\}$ in the gradient descent way, i.e. $\Delta\theta_i = -\lambda_0 \partial E^r / \partial \theta_i$.

Step 4: Using a training loop between the local left hemisphere and the local right hemisphere to minimize the cost function in the gradient descent way, i.e. $\Delta\theta_i = -\lambda_0 \partial E / \partial \theta_i$.

Step 5: If the cost function is smaller than a given threshold, then uploading the updated parameters of the local right hemisphere to the core machine. Otherwise, return to the step 2.

Step 6: Collecting the updated parameters from the terminal machines. If there is a new label, then using a training loop between the global left hemisphere and the global right hemisphere to minimize the cost function in the gradient descent way, i.e. $\Delta\theta_i = -\lambda_0 \partial E / \partial \theta_i$. Otherwise, using the updated parameters to minimize the training error of the global left hemisphere $B_{CL}$ in the gradient descent way, i.e. $\Delta\theta_i = -\lambda_0 \partial E^t / \partial \theta_i$.

Step 7: If the training error of the global left hemisphere $B_{CL}$ is smaller than a given threshold, then exchanging the parameters with the local left hemispheres $\{B^{(i)}_{EL}\}$, and ending the procedure. Otherwise, return to the step 6.

The pre-training procedure to construct a unidirectional neural network based on the equation (1) is as the following.

Procedure (II): The construction of the unidirectional neural network.
Input: A unidirectional full-connected layer-wise neural network with the set of parameters $\theta$ and the set of centralized data $D_c$.
Output: A unidirectional neural network with the set of pre-trained parameters.
Step 1: Initiating the parameters.
Step 2: Adjusting the parameters in the gradient descent way with respect to the training error $E^t$, i.e. $\Delta\theta_i = -\lambda_0 \partial E^t / \partial \theta_i$.
Step 3: If the current parameter is smaller than a given threshold, calculating the training error difference $\Delta E^t$ with and without the current parameter. Deleting the current parameter with the probability of $min[1, exp(-\Delta E^t / E^t)]$.
Step 4: Judging that the degrees of all neurons are smaller than $d_{max}$. If the answer is NO, returning to step 2. If the answer is YES, ending the procedure.

The pre-training procedure to construct a bidirectional neural network based on the equation (2) is as the following.

Procedure (III): The construction of the bidirectional neural network.
Input: A bidirectional full-connected layer-wise neural network with the set of parameters $\theta$ and the set of centralized data $D_c$.
Output: A bidirectional neural network with the set of pre-trained parameters.
Step 1: Initiating the layer number ($h=0$).
Step 2: Initiating the parameters between the $h^{th}$ layer and the $(h+1)^{th}$ layer.
Step 3: Adjusting the parameters between the $h^{th}$ layer and the $(h+1)^{th}$ layer in the gradient descent way with respect to the generation error $E^r$, i.e. $\Delta\theta_i = -\lambda_0 \partial E^r / \partial \theta_i$, where $\lambda_0$ is a user-defined non-zero small number.
Step 4: If the current parameter is smaller than a given threshold, calculating the generation error difference $\Delta E^r$ with and without the current parameter. Deleting the current parameter with the probability of $min[1, exp(-\Delta E^r / E^r)]$.
Step 5: Judging that the degrees of all neurons on the $h^{th}$ layer are smaller than $d_{max}$. If the answer is NO, returning to step 3. If the answer is YES, shifting to the next step.
Step 6: Judging the sum of all degrees of the $h^{th}$ layer in the positive direction from $\{x_i\}$ to $\{y_j\}$ is smaller than that of the $(h-1)^{th}$ layer or not. If the answer is NO, returning to step 3. If the answer is YES, shifting to the next step.
Step 7: Judging that the current $h$ is larger than $H$ or not. If the answer is NO, returning to step 2, and let $h$ be $h+1$. If the answer is YES, ending the procedure.

## 6. Evaluation

Python 2.7.11 [16] and Theano 0.8 [17] are used to evaluate the proposed dual-hemisphere model of machine learning in comparison to a state-of-the-art models of deep feed forward neural networks (DFFNN). Both models were implemented on a computer cluster of a core machine and $M$ terminal machines connected by an Ethernet switch with a uniform distribution of transmission delay. The core machine had 24 processors of 2.10GHz Intel Xeon E5-2620 CPU 32GB RAM and NVIDIA GPU Grid K2 8GB GDDR5. Each terminal machine had 2 processors of 2.50GHz Intel Core i7-4710MQ CPU 16GB RAM. The communication cost between the terminal machines and the end users was set to be zero.

In experiment, the initial setting of the bidirectional neural network and the unidirectional neural network were as follows. The input layer was set to be 96 neurons. The sigmoid output layer had 8 neurons with an additional softmax output layer for responses to the user. Each number of neurons in a given hidden layer was set to be a layer-wise decreasing number as follows. The first hidden layer had 1000 neurons. The second hidden layer had 900 neurons. The third hidden layer had 800 neurons. The fourth hidden layer had 700 neurons. The fifth hidden layer had 600 neurons. The weights ($\lambda_1$ and $\lambda_2$) of regularization terms ($R_1$ and $R_2$) were set to be two small non-zero numbers, where $\lambda_1 = 0.00001$ and $\lambda_2 = 0.00009$ were two empirical values, to avoid the over-fitting of machine learning.

In an application on the multimedia service, we use $M = 3, 4, 5, 6, 7$ to test the performance and the default value of $M$ is 5. Each terminal machine simulated an end-user and a part of the multimedia service system. The measured data include the timestamp, the latency, the video quality, the downloading speed and the video source. The gathered data could include the user behavior, the user feedback and the watching history. The video sources include but are not limited to the user-generated video server, the caching server and the online web sites such as Youku [18], Tudou [19], QQ Video [20], Sohu [21], CNTV [22] and Letv [23]. The request from a user is to find the best video source with the widest bandwidth. The input is the past bandwidths calculated by the measured downloading speeds for different video sources. The sigmoid output is the predicted bandwidths for different video sources. The softmax output is one of the video sources with the widest bandwidth. Then, the response to the user is to recommend the best video source according to the softmax output.

The accuracy was defined as $N_c/N_t$, where $N_c$ was the number of the correct responses to recommend the best video source within a given limit of latency and $N_t$ is the total number of requests. The total latency included the transmission delay and the processing time. If the latency was longer than the given limit or the bandwidth was not the widest one according to the measurement, the response was regarded to be incorrect. The measurement and the preprocessing were conducted from December 15, 2015 to December 23, 2015 using the existing methods [24]-[25]. We totally collected 210,011 samples. All samples were divided into three data sets according to the timestamp. The 80% samples were used to adjust the model parameters for training. The 10% samples were used to evaluate the cost function for validation. The remaining 10% samples were used to test the accuracy. The accuracy was evaluated by recommending a video source with a short latency and the widest bandwidth in comparison to the measured data. Besides, the samples was repeated with $p$ times to evaluate the scalability performance, where $p = 1, 2, 4, 8, ..., 1024$.

**Table 1** The accuracy with respect to the value of $M$ without the limitation of latency.

| The value of $M$ | Accuracy (%) DFFNN | Accuracy (%) Proposal | Improvement (%) |
|---|---|---|---|
| 3 | 87.46 | 92.06 | 5.25 |
| 4 | 87.46 | 92.73 | 6.02 |
| 5 | 87.46 | 93.04 | 6.37 |
| 6 | 87.46 | 93.27 | 6.64 |
| 7 | 87.46 | 93.29 | 6.66 |

**Table 2** The accuracy with respect to the latency based on the comparison between prediction and measurement.

| Latency (s) | Accuracy (%) DFFNN | Accuracy (%) Proposal | Improvement (%) |
|---|---|---|---|
| 0.1 | 42.37 | 93.46 | 51.09 |
| 0.2 | 50.95 | 93.90 | 42.95 |
| 0.3 | 56.13 | 93.47 | 37.33 |
| 0.4 | 61.57 | 92.54 | 30.97 |
| 0.5 | 67.95 | 93.39 | 25.44 |
| 0.6 | 72.19 | 94.43 | 22.24 |
| 0.7 | 77.04 | 93.33 | 16.30 |
| 0.8 | 81.35 | 91.47 | 10.12 |
| 0.9 | 85.92 | 93.96 | 8.04 |
| 1.0 | 87.39 | 93.81 | 6.42 |

To evaluate the communication cost, the cluster was described as a graph $(V, E, ED, EF)$, where $V$ was the vertex set $\{i\}$, $E$ was the edge set $\{e_{ij}\}$, $ED$ was the set of edge geographical distance $\{ed_{ij}\}$, and $EF$ was the set of edge flow $\{ef_{ij}\}$ during communication. The communication cost was defined as $CC = \sum_{ij} ed_{ij}\, ef_{ij}$. For the tree topology of computer cluster of a core machine and $M$ terminal machines, the communication cost was simplified as $CC_{tree} = \sum_{i} ed_i\, ef_i$, where $ed_i$ was the geographical distance between the core machine and the terminal machines, $ef_i$ was the data flow between the core machine and the terminal machines using model communication, and $M$ was the number of the terminal machines.

We evaluated a state-of-the-art model of deep feed forward neural network (DFFNN) [26]-[27] which was extended to be an application to the online multimedia service. DFFNN was trained by a back-propagation to get the optimized performance. The average accuracy of the DFFNN model without the limitation of latency is shown in Table 1. There is no difference for the centralized DFFNN model when all distributed data are collected in the data center. The accuracy of the DFFNN model with respect to the latency was shown in Table 2. We could see that the accuracy was low when the requirement of the latency was short. The communication cost of the DFFNN model was shown in Table 3. We could see that the communication was expensive with respect to the scaling of data size.

**Table 3** The improvement of the communication cost with respect to the number of samples.

| Samples (100K) | Communication Cost (Mb) DFFNN | Communication Cost (Mb) Proposal | Improvement (times) |
|---|---|---|---|
| 1 | 7.38 | 1.78 | 4.14 |
| 2 | 13.99 | 2.43 | 5.76 |
| 4 | 28.07 | 4.97 | 5.65 |
| 8 | 55.95 | 9.74 | 5.75 |
| 16 | 111.62 | 18.91 | 5.90 |
| 32 | 222.65 | 37.75 | 5.90 |
| 64 | 445.55 | 74.66 | 5.97 |
| 128 | 890.27 | 149.41 | 5.96 |
| 256 | 1780.76 | 299.83 | 5.94 |
| 512 | 3560.97 | 595.52 | 5.98 |
| 1024 | 7120.97 | 1190.78 | 5.98 |

**Table 4** The comparison of the learning time between a state-of-the-art model and the proposed terminal machine.

| Samples (100K) | Learning Time (s) DFFNN | Learning Time (s) Terminal | Cost (times) |
|---|---|---|---|
| 1 | 12.25 | 62.25 | 5.08 |
| 2 | 23.63 | 116.78 | 4.94 |
| 4 | 45.37 | 237.45 | 5.23 |
| 8 | 91.41 | 473.96 | 5.19 |
| 16 | 180.89 | 874.41 | 4.83 |
| 32 | 374.55 | 1788.47 | 4.77 |
| 64 | 725.93 | 3761.53 | 5.18 |
| 128 | 1474.34 | 7049.08 | 4.78 |
| 256 | 2960.77 | 15825.06 | 5.34 |
| 512 | 5858.70 | 29243.68 | 4.99 |
| 1024 | 11737.34 | 63966.82 | 5.45 |

We evaluated the proposed model based on the wide area network intelligence. It was constructed and trained according to the procedure (I), (II) and (III). If the limit of latency is long enough, the final response could be the long-distance response. Otherwise, it is set to be the terminal response. The average accuracy of the proposed model with respect to the value of $M$ without the limitation of latency is shown in Table 1. We could see that the improvement of accuracy was scalable and stable. The accuracy and the latency of the proposed terminal models was shown in Table 2. We could see that the accuracy was considerably improved when the latency requirement was short. The communication cost of the proposal was shown in Table 3. We could see that the communication was stably improved with respect to the scaling of samples in comparison to the state-of-the-art DFFNN model.

**Table 5** The comparison of the learning time between a state-of-the-art model and the proposed core machine.

| Samples (100K) | Learning Time (s) DFFNN | Learning Time (s) Core | Cost (times) |
|---|---|---|---|
| 1 | 12.25 | 33.02 | 2.70 |
| 2 | 23.63 | 63.47 | 2.69 |
| 4 | 45.37 | 123.22 | 2.72 |
| 8 | 91.41 | 267.05 | 2.92 |
| 16 | 180.89 | 371.74 | 2.06 |
| 32 | 374.55 | 803.71 | 2.15 |
| 64 | 725.93 | 1626.05 | 2.24 |
| 128 | 1474.34 | 3500.64 | 2.37 |
| 256 | 2960.77 | 8856.20 | 2.99 |
| 512 | 5858.70 | 14578.57 | 2.49 |
| 1024 | 11737.34 | 30209.03 | 2.57 |

The cost of above improvement was longer learning time with respect to the number of training samples. As shown in Table 4, the learning time of the state-of-the-art DFFNN model was shorter than that of the proposed model in the terminal machine.

Furthermore, as shown in Table 5, the learning time of the proposed model in the core machine was a little longer than that of the state-of-the-art DFFNN model.

## 7. Discussion

As mentioned above, many state-of-the-art models for multimedia service and recommendation have been investigated. T. Ghalut *et al.* [26] proposed a feed-forward random neural network to classify video content according to the impact of video content on video quality. A. Tomar *et al.* [27] introduced an approach for hashtag recommendation using deep feed-forward neural network and a skip-gram model to learn distributed word representations.

However, the mentioned techniques are applicable to data center or local area network. It is hard to be used in wide area network due to the challenges of long-distance communication and latency. In fact, most network data from the end users are geo-distributed. This makes cloud computing inefficient. The novelty of the proposal lies two hemispheres with different functions. The left one is for on-site response, while the right one is for data generation. Furthermore, a loop training between left and right hemispheres is designed for parameter update without data storage. It is important because the volume of data storage is quite limited for terminals.

As a result, the proposal improved the performance of communication and latency. The proposal not only makes it easy to provide real-time service for end-users but also makes the communication secure and private because there is no original data to communicate between the core machine and the terminal machines. The learning time is the cost of above improvement, but it is a relatively trivial performance in case of the increasingly large volume of network data [28-29].

The reason why the proposal achieves higher accuracy than the centralized DFFNN model is because the geo-distributed data are personalized in each terminal machine. Even if all of the distributed data are collected in the data center, it is still hard to improve the accuracy of the DFFNN model due to the loss of the personalized features in different geographical sites. Furthermore, the reason why the proposal achieves shorter latency is because the response of the proposed model is directly from one of the terminal machines in the neighborhood of the end users, instead of the data center for the centralized models.

## 8. Conclusion

In this paper, a dual-hemisphere model of machine learning has been proposed for wide area network intelligence. The architecture of machine learning was modified to improve the latency by terminal response and the communication by data generation. The application on multimedia service showed that the proposal performs better than a state-of-the-art model of deep feed-forward neural network in terms of communication, accuracy and latency. Evaluations also showed the cost of above improvement was longer learning time.

**Acknowledgment**

This work is partly supported by the Special Fund for Strategic Pilot Technology of Chinese Academy of Sciences under Grant No. XDA06040501. The authors would like to thank the anonymous reviewers for their valuable comments and suggestions.